\documentclass[12pt]{article}

\usepackage{amsmath}
\usepackage{graphicx}
\usepackage{cite}

\newcommand{\mys}[1]{\section{#1}
	\setcounter{equation}{0}}
	\renewcommand{\theequation}{\arabic{section}.\arabic{equation}}
\newcommand{\myappendix}{\appendix
   \renewcommand{\theequation}{\Alph{section}.\arabic{equation}}
   \vspace{30pt} \noindent {\Large \bf Appendices}}

\newlength{\dummysp}
\newcommand{\half}{{\frac{1}{2}}}

\newcommand{\beq}{\begin{eqnarray}}
\newcommand{\eeq}{\end{eqnarray}}
\newcommand{\nnn}{ \nonumber \\ }

\newcommand{\Kcal}{{\cal K}}

\newcommand{\Zbf}{{{\bf Z}}}
\newcommand{\Rbf}{{{\bf R}}}

\newcommand{\chib}{{\bar \chi}}
\newcommand{\e}{{\epsilon}}
\newcommand{\s}{{\sigma}}

\newcommand{\ord}[1]{{{\cal O}(#1)}}
\newcommand{\gappeq}{\mathrel{\rlap {\raise.5ex\hbox{$>$}}
{\lower.5ex\hbox{$\sim$}}}}
\newcommand{\lappeq}{\mathrel{\rlap{\raise.5ex\hbox{$<$}}
{\lower.5ex\hbox{$\sim$}}}}
\newcommand{\myref}[1]{(\ref{#1})}

\newcommand{\bfe}[1]{\vspace{5pt} {\bf #1 \hspace{2pt}}}
\newcommand{\ite}[1]{\vspace{5pt} {\it #1 \hspace{2pt}}}

\newcommand{\ben}{\begin{enumerate}}
\newcommand{\een}{\end{enumerate}}

\newcommand{\bit}{\begin{itemize}}
\newcommand{\eit}{\end{itemize}}
\newcommand{\Lcal}{{\cal L}}

\newcommand{\mubar}{{\bar \mu}}

\newcommand{\muhat}{{\hat \mu}}
\newcommand{\nuhat}{{\hat \nu}}

\newcommand{\Bcal}{{\cal B}}

\newcommand{\Bcala}{{\Bcal_a}}

\newcommand{\Bcalaa}{{\Bcal_{2a}}}
\newcommand{\chit}{{\tilde \chi}}

\newcommand{\Atil}{{\tilde A}}
\newcommand{\RL}{Reisz-L\"uscher}
\newcommand{\delb}{{\bar \delta}}

\newcommand{\msg}[2]{(\overline{\overline{#1 \otimes #2}})}
\newcommand{\cReisz}{\cite{Reisz:1987da}}
\newcommand{\pia}{\frac{\pi}{a}}

\newcommand{\bnb}{{\Big|\Big|}}
\newcommand{\delba}{{\delb_a}}
\newcommand{\deltw}{\delta^{[2]}}
\newcommand{\modtxt}{\text{mod}~}
\newcommand{\pibf}{{\boldsymbol{\pi}}}
\newcommand{\tpia}{{\frac{2\pi}{a}}}
\newcommand{\degr}[1]{\overline{\hbox{degr}}_{\hat {#1}}}
\newcommand{\ddegr}[1]{\overline{\overline{\hbox{degr}}}_{\hat {#1}}}
\newcommand{\PCThm}{power-counting theorem}

\def\[{\left [}
\def\]{\right ]}
\def\({\left (}
\def\){\right )}

\begin{document}
\begin{titlepage}

\renewcommand{\thefootnote}{\fnsymbol{footnote}}

FTPI-MINN-06/19
\hfill June 26, 2006

UMN-TH-2506/06
\hfill hep-lat/0606003

\vspace{0.45in}

\begin{center}
{\bf \Large Power-counting theorem \\ \vskip 10pt
for staggered fermions}
\end{center}

\vspace{0.15in}

\begin{center}
{\bf \large Joel Giedt\footnote{{\tt giedt@physics.umn.edu}}}
\end{center}

\vspace{0.15in}

\begin{center}
{\it Fine Theoretical Physics Institute, University of Minnesota \\
116 Church St.~S.E., Minneapolis, MN 55455 USA }
\end{center}

\vspace{0.15in}

\begin{abstract}
Lattice power-counting is extended to QCD with
staggered fermions.  As preparation,
the difficulties encountered by Reisz's original
formulation of the lattice power-counting theorem
are illustrated.
One of the assumptions that is used
in his proof does not hold for staggered
fermions, as was pointed out long ago
by L\"uscher.  Finally, I generalize the power-counting
theorem, and the methods of Reisz's proof,
such that the difficulties posed by staggered fermions
are overcome.
\end{abstract}

\end{titlepage}

\renewcommand{\thefootnote}{\arabic{footnote}}
\setcounter{footnote}{0}


\mys{Motivation and summary}
Lattice QCD with improved staggered fermions (SFs),
or, Kogut-Susskind fermions \cite{Kogut:1974ag,Susskind:1976jm,
Sharatchandra:1981si}, 
has recently enjoyed publicity
for its ability to correctly reproduce many aspects of
hadronic physics with reasonable 
accuracy \cite{Davies:2003ik,ptart}.  
However, SFs have some notable properties.  For instance,
SFs do not entirely overcome the fermion 
doubling problem.  Rather, they reduce the
number of continuum modes from 16 to 4.  (A
further reduction to 2 modes is possible,
by projecting quarks and antiquarks to
odd and even sublattices resp.~\cite{Sharatchandra:1981si}.)
These 4 modes are referred to as {\it tastes,}
to distinguish them from the $N_f$ flavors
in the continuum theory.  To
estimate the fermion measure of $N_f$
continuum flavors, one takes the power $N_f/4$ of the
fermion determinant in the definition of
the functional integral.  I will not address
the attendant controversy, but rather another
technical question:  lattice power-counting for
staggered fermions.
Here again, fermion doubling creates difficulties,
as will be discussed at some length in this article,
and as was pointed out some years 
ago by L\"uscher \cite{Luscher:1988sd}.


To better understand perturbative renormalization of SFs
it is of course
useful to have a lattice power-counting theorem.
By way of analogy, renormalizability of $SU(N)$ Yang-Mills
coupled to Wilson fermions has been proven
some time ago by Reisz \cite{Reisz:1988kk}.  This result
was based on his earlier work on BPHZ-like renormalization
theory on the lattice \cite{Reisz:1987px,Reisz:1987hx}.
That work rested crucially on his
lattice power-counting theorem
\cite{Reisz:1987da,Reisz:1987pw}.
(This literature is rather mathematical; more accessible
reviews are those by Reisz \cite{Reisz:1987at,Reisz:1988zv} 
and L\"uscher \cite{Luscher:1988sd}.)
Reisz's lattice power-counting theorem
was a significant achievement because
on the lattice Feynman integrands are trigonometric
rather than rational functions of momenta;
this can lead to results that differ from
those of the continuum in important ways.\footnote{
An amusing example occurs, for instance, in naive discretizations
of supersymmetric quantum mechanics \cite{Giedt:2004vb}.}

It is often stated that no power-counting theorem exists
for SFs; for example 
in Refs.~\cite{Jansen:2003nt,Pernici:1995ij,Adams:2004wp,Lel06}.
However, it is also widely believed that the theory of
SFs coupled to Yang-Mills (denoted here SF-QCD) yields the
right quantum continuum limit in perturbation
theory.  That is to say, the lattice perturbation
series can be renormalized
and matched to a continuum renormalization scheme at every
order in the gauge coupling $g$.  
This conclusion is supported by an analysis of
the types of non-irrelevant operators that
are allowed by the symmetries of SF-QCD.  One
finds that all such operators are already
present at tree-level.  (See for example
\cite{Golterman:1984cy} and refs.~therein.)
That is, from a Wilsonian point of view one
concludes that SF-QCD is in the same universality
class as continuum QCD.  It is reasonable to
believe that by an adjustment of the bare parameters of the
lattice action, one can arbitrarily adjust the
coefficients of all non-irrelevant operators
in the infrared, in order to obtain the desired theory.

The belief that SF-QCD is renormalizable
also follows from a consideration of
powers of the lattice spacing $a$ that
arise in vertices and propagators of the theory,
and how they appear in loop diagrams, an early example being
\cite{Sharatchandra:1981si}.  In fact,
for 1-loop diagrams, it is easy to power-count
by partitioning the loop integration domain
in a sensible way and estimating the integrand
and measure for each of those domains.
But this is nothing
other than a limited version lattice power-counting.  So, in fact,
a version of power-counting already exists, though
it is not as general as we would like.
In actuality, this sort of partitioning is
exactly what is done in Reisz's proof of
his lattice power-counting theorem.  However,
the complexities that occur at high orders---where the
number of domains increases factorially---are
best addressed by a more sophisticated
mathematical approach, just as in the continuum
proofs of Weinberg \cite{Weinberg:1959nj} or
Hahn and Zimmermann \cite{HZ:68}.
It is this sort of general method of power-counting
that is aimed at in the present study.

I now summarize the remainder of this article:
\bit
\item
In \S\ref{s:bases}, I briefly review two well-known
formulations of the SF-QCD action, and the
corresponding Feynman rules.
\item
In \S\ref{s:pcrv}, I review the conditions
for the Reisz power-counting theorem.  I also
remind the reader of the lattice UV degree 
(of divergence) that is defined in Reisz's theorem.
\item
In \S\ref{s:pca} the conditions of the
Reisz theorem are examined for the two
formulations of SF-QCD thate were described in \S\ref{s:bases}.
It is shown that in both cases the conditions of
Reisz's theorem are violated.
I explain the essential, basis-independent
reason for this failure.
\item
In \S\ref{rcpg} I generalize Reisz's theorem
and methods of proof
in such a way that lattice power-counting
can be applied to staggered fermions.
\item
In \S\ref{s:con} I conclude
with a summary and discussion of further issues that
could be explored.  
\eit
Various appendices are included for details
that would detract from the main discussion,
but that are essential to the proof:
\bit
\item
In \S\ref{rid1} I discuss resolutions of identity
that are used in a domain decomposition for the
loop momenta integration.  Both the one used by
Reisz, and a generalized one that is applied
in \S\ref{rcpg} are given.
\item
In \S\ref{smsr} I discuss a simplification
of the Feynman rules that is very useful
in the momentum-space taste basis (MSTB).
\item
In \S\ref{ddt}, it is shown that the domain
of internal momenta can be extended in a useful
way in the MSTB.
\eit

\mys{Bases}
\label{s:bases}
\subsection{The 1-component basis}
\label{1cbd}
The gauge covariant SF action with link fields $U_\mu(r)$
is just \cite{Kogut:1974ag,Susskind:1976jm,Sharatchandra:1981si}
\beq
S_{SF} &=& \sum_{r\in \Zbf^4} \sum_{\mu=1}^4 \half a^3 \alpha_\mu(r)
\[ \chib(r) U_\mu(r) \chi(r + \muhat) 
- \chib(r + \muhat) U_\mu^\dagger(r) \chi(r) \] \nnn
&& + \; \sum_{r\in \Zbf^4} m a^4 \chib(r) \chi(r),
\qquad \alpha_\mu(r) \equiv (-)^{r_1 + \cdots + r_{\mu-1}}.
\label{sfacti}
\eeq
Color indices are suppressed, $r$ is a site index,
and $a$ is the lattice spacing.
I refer to this as the {\it 1-component basis (1CB).}
Under the lattice translation $\phi(r) \to \phi(r+s)$,
with $\phi = U_\mu,\chi,\chib$, the action is
only invariant for even shifts $s \in 2\Zbf^4$.
Thus what I will call the {\it K\"ahler-Dirac lattice}\footnote{This
is the lattice generated by the basis vectors $2\muhat$.
I.e., those corresponding to translations
that leave the lattice action invariant.
It is the lattice through which free SFs
are related to K\"ahler-Dirac fermions \cite{Rabin:1981qj,
Banks:1982iq,Becher:1981cb,Becher:1982ud,Becher:1982xa}.}
consists of elementary cells that are hypercubes
of length $b \equiv 2a$ on each side, denoted $2a\Zbf^4$.

The free fermion propagator ($U_\mu \equiv 1$) has 16 lattice
poles; i.e., minimal eigenvalues
of the (Euclidean) SF Dirac operator.  The
16-fold degeneracy corresponds to 4 continuum 
Dirac fermions.  In momentum space,
the additional poles lie at edges of the first
Brillouin zone 
\beq
\Bcala=(-\pi/a,\pi/a]^4.
\label{1BZ}
\eeq
More specifically, the poles lie
at the sites of the lattice $(\pi/a) \Zbf^4$ that is
reciprocal to the K\"ahler-Dirac lattice $2a\Zbf^4$.

Perturbation theory is, as usual, defined by expansion of
\beq
U_\mu(r) \equiv \exp iag A_\mu(r)
\label{Uexp}
\eeq
in powers of $g$.  Fourier transforms are defined with conventions:
\beq
\phi(r) &=& \int_\Bcala \frac{d^4k}{(2\pi)^4} e^{iar \cdot k} \tilde \phi(k),
\quad \phi \in \{ \chi, \chib, A_\mu \},
\label{1cft}
\eeq
where $\Bcala$ is defined in \myref{1BZ}.
I have chosen to make $k$ dimensionful, since
it agrees with the conventions of Reisz.
It will prove useful below to periodically
extend the fields:  
\beq
\tilde \phi(k + (2\pi/a)z) \equiv \tilde \phi(k), \quad
\forall \quad z \in \Zbf^4.
\label{pere}
\eeq
The form of $\Atil_\mu(k)$
follows the convention of \cite{Golterman:1984cy} but
differs from the convention of \cite{Daniel:1987aa,Patel:1992vu}
by a factor of $e^{iak_\mu/2}$.  The choice that is
made here gives Feynman rules that are manifestly
periodic on $\Bcal_a$ since $\tilde \phi(k+(2\pi/a)\nuhat) = 
\tilde \phi(k) \; \forall \nu \in \{ 1, \ldots, 4\}$, 
and $\tilde \phi \in \{ \chit, \tilde \chib,
\Atil_\mu \}$.  In the conventions of \cite{Daniel:1987aa,Patel:1992vu},
slightly more effort must be expired
to demonstrate periodicity
of numerators of Feynman diagrams on $\Bcal_a$,
since individual vertex factors lack
this property.  $2\pi/a$-periodicity of the
numerator of Feynman integrands
is an important assumption in Reisz's proof.
See for example the \RL\ conditions
{\bf V1} and {\bf C1} in \S\ref{s:pcrv}.

For instance, in the conventions of \cite{Daniel:1987aa,Patel:1992vu}
the $\ord{g}$ gluon-quark vertex is proportional to
$\cos(p_\mu a + \half k_\mu a)$, with $p$
incoming momentum on the $\tilde \chib$ line
and $k$ momentum on the incoming gluon
line.  Suppose $k$ is a loop
momentum and we want to check the
$2\pi/a$-periodicity in condition {\bf V1.}
Under $k \to k + (2\pi/a)\muhat$ the
vertex reverses sign.  This corresponds to
$\Atil_\mu(k+(\pi/a)\muhat) = -\Atil_\mu(k)$,
due to the additional factor of $e^{iak_\mu/2}$ in
the Fourier transform of \cite{Patel:1992vu,Daniel:1987aa}.  
However, one finds that
the sign is always cancelled in some other
part of the diagram that also involves the
loop momentum $k$.  This of course must
be true, due to the equivalence with
the formulation that I choose, where the
Feynman rules themselves enjoy periodicity
on $\Bcala$.

The propagators and leading boson-fermion vertices are given,
for instance, in Table 1 of \cite{Daniel:1987aa}, apart
from the factor $e^{-ik_\mu a/2}$
for each incoming $\Atil_\mu(k)$.
Let $\delb_a$ denote the $2\pi/a$-periodic $\delta$-function.
Also define $\mubar = \sum_{\nu=1}^{\mu-1} \nuhat$.  Thus 
\beq
\bar 1 = (0,0,0,0), \quad \bar 2 = (1,0,0,0), \quad
\bar 3 = (1,1,0,0), \quad \bar 4 = (1,1,1,0).
\eeq
These are useful
because in \myref{sfacti} we can write
$\alpha_\mu(r) = \exp i \pi \mubar \cdot r$,
which makes the Fourier transform easy to compute.

Ignoring ghosts and pure YM vertices,\footnote{There is
no difficulty applying Reisz's \PCThm\ to pure YM.
It is the SF propagator and vertices that pose
problems.  I include the gluon propagator (in Feynman
gauge) for the purpose of illustrating how
the YM sector is treated in tandem with SFs
when PC is attempted.  Treatment of
ghosts is identical.  A detailed analysis of
pure YM interactions will not be required
in what follows.} the Feynman rules are:
\beq
D_{\mu \nu}^{cd}(k,\ell) &=& \delta^{cd} \delta_{\mu \nu} \delb_a(k+\ell)
\[ \sum_\mu \frac{4}{a^2} \sin^2 \frac{k_\mu a}{2} + \lambda^2 \]^{-1},
\\
S^{ij}(p,q) &=& \delta^{ij} \frac{ m \delb_a(p+q)
- ia^{-1} \sum_\mu \sin (p_\mu a)
\delb_a(q + p+ \frac{\pi}{a} \mubar) }
{a^{-2} \sum_\mu \sin^2 (p_\mu a) + m^2},
\label{fp1b} \\
V_\mu^{c;ij}(p,q;k) &=& -\frac{i}{2} g(T_R^c)^{ij}
\( e^{i p_\mu a} + e^{-i(p_\mu +k_\mu) a} \)
\delb_a(k+p+q+\frac{\pi}{a} \mubar).
\label{1dpr}
\eeq
$T_R^c$ are generators of the gauge group
in the quark representation $R$;
$i,j$ are color indices;
$\lambda$ is an IR regulating mass
for the gluon.
$p,q$ are incoming momenta on the $\tilde \chib, \chit$
lines respectively, whereas $k,\ell$ are incoming
momenta on $\Atil_\mu$ lines.  Note that momentum conservation
is only mod $\pi/a$ where the fermions are concerned
(i.e., one finds $(\pi/a)\mubar$ inside the $\delba$-functions),
due to the fact that the K\"ahler-Dirac lattice is $2a\Zbf^4$
(in physical units), which has for a reciprocal lattice
$(\pi/a)\Zbf^4$.

\subsection{Momentum space taste basis}
\label{mstb}
Here I discuss the momentum space taste basis (MSTB)
that was originally introduced
in \cite{Sharatchandra:1981si}.  I present the
results in conventions that are similar to
\cite{Daniel:1987aa,Patel:1992vu}; I retain
the modification of the gluon Fourier
transform that was discussed above.\footnote{There
is also a position space taste basis \cite{Gliozzi:1982ib},
which I will not discuss here, except briefly in \S\ref{mism}.  
For a more detailed exposition, consult also 
\cite{Kluberg-Stern:1983dg} as well
as the reviews \cite{Rothe:2005nw,Montvay}.}

We make the following redefinition of the momentum 
space 1-component fields:
\beq
&& \tilde \chi(k) = \chi_A(k_r), \quad
\tilde \chib(k) = \chib_A(k_r), \quad
k=k_r+ \pia A \mod \tpia,
\nnn
&& k_r = \pibf(k) \in \Bcalaa \equiv 
( -\pi / 2a, \pi/2a ]^4, \nnn
&& A \in \Kcal, \quad 
\Kcal \equiv \{ (0^4), (\underline{1,0^3}),
(\underline{1^2,0^2}), (\underline{1^3,0}), (1^4) \} .
\label{rbaa}
\eeq
The notation is as follows.  In the definition
of the set of 4-vectors $\Kcal$, powers indicate how many
times a 0 or 1 appears.  Underlining indicates that
all permutations of entries are to be included.
Note that the 16 lattice poles described in \S\ref{1cbd} above
are located in momentum space at $k \in (\pi/a) \Kcal$.
The map $\pibf$ is a projection to the {\it reduced}
first Brillouin zone $\Bcalaa$.
Feynman vertices and propagators involving the
fermions are then translated from the 1CB
using this identification.  In practice it is
helpful to extend the definition as follows:
\beq
\tilde \chi(k) = \chi_A(k'), \quad
\tilde \chib(k) = \chib_A(k'), \quad
k=k' + \pia A, \quad \forall \quad k,k' \in \Rbf^4,
~ A \in \Kcal.
\label{kedf}
\eeq
Here, the periodically extended definitions [cf.~\myref{pere}]
of $\tilde \chi(k), \tilde \chib(k)$ are used.

Taking these redefinitions into account, corresponding
to \myref{fp1b} and \myref{1dpr} we have the Feynman
rules \cite{Daniel:1987aa,Patel:1992vu}:
\beq
&& S^{ij}\(p = p' + \pia A, q = q' + \pia B \) 
\equiv S_{AB}^{ij}(p',q') \nnn
&& \quad = \delta^{ij} \delb_a(p'+q') 
\frac{ m \msg{1}{1}_{A,B}
- ia^{-1} \sum_\mu \sin (p'_{\mu} a) \msg{\gamma_\mu}{1}_{A,B}}
{a^{-2} \sum_\mu \sin^2 (p'_{\mu} a) + m^2},
\label{msfp} \\
&& V_\mu^{c;ij} \(p=p'+ \pia A, q=q'+ \pia B; k \) 
\equiv V_{\mu;AB}^{c;ij}(p',q';k) \nnn
&& = -\frac{i}{2} g (T_R^c)^{ij} \delb_a(p'+q'+k)
\( e^{i p'_{\mu} a} + e^{-i(p'_{\mu} + k_\mu) a} \)
\msg{\gamma_\mu}{1}_{A,B} . \qquad
\label{ms1v}
\eeq
As in \myref{kedf}, $p',q',k$ take any values in $\Rbf^4$.
An equivalence has been used to obtain momentum
conserving (mod $2\pi/a$) $\delb_a$-functions.
It is reviewed in \S\ref{smsr} and plays
a crucial role in the generalized proof of \S\ref{rcpg}.
The $16 \times 16$ momentum space spin-taste
matrices $\msg{1}{1}_{A,B}$ and $\msg{\gamma_\mu}{1}_{A,B}$
are written in the notation of \cite{Daniel:1987aa,Patel:1992vu}.
Definitions can be found therein; we will not
need their explicit form in what follows.

The lattice perturbation theory also contains quark-multigluon
vertices that are irrelevant operators, suppressed
by explicit powers of the lattice spacing.
They are important to take into account for
the cancellation of divergences (see for example
\S14.2 of \cite{Rothe:2005nw}), and are easily incorporated
into the formalism that is discussed below.\footnote{
I thank David Adams for an important discussion
regarding the role of the irrelevant vertices.}

\mys{Review of Reisz lattice power-counting}
\label{s:pcrv}
\subsection{The lattice UV degree}
A given Feynman integral is written in the
general form
\beq
\hat I = \int_{\Bcala^L} d^4k_1 \cdots d^4k_L
~ \frac{V(k,q;m,a)}{C(k,q;m,a)}.
\label{fdgf}
\eeq
Here, $k_1,\ldots,k_L$ are loop momenta and $q_1,\ldots,q_E$ 
are external momenta.  Note that the loop momenta
are integrated over $\Bcala^L = \times^L \Bcala
= (-\pi/a,\pi/a]^{4L}$; cf.~\myref{1BZ}.
Also, $m$ stands collectively for mass parameters.  The numerator
$V$ incorporates vertex factors and the numerators of
propagators; $C$ consists of a product of
propagator denominators.

Reisz defines the UV degree of $V$ and $C$,
and thence of the integral $\hat I$.
At higher orders, this must be done over
Zimmermann subspaces $H$.  To each propagator
corresponds a {\it line momentum} $\ell_i(k,q)$
(cf.~\myref{Lmdf} below).
There is a sense in which $L$ of these form a basis
w.r.t.~$k_1,\ldots,k_L$, as will be
explained in \S\ref{RLc} below (cf.~condition
{\bf L2}).  We decompose this set as follows:
\beq
&& u_1 = \ell_{i_1}, \quad \ldots, \quad u_d=\ell_{i_d}; \nnn
&& v_1 = \ell_{j_1}, \quad \ldots, \quad v_{L-d}=\ell_{j_{L-d}}.
\label{uvdf}
\eeq
We regard $v_1,\ldots,v_{L-d}$ and $q_1,\ldots,q_E$ as
external momenta to the Zimmermann subspace $H$.
The $u_1,\ldots,u_d$ are regarded as internal momenta that
provide a parameterization of $H$.  We denote the
set of all Zimmermann subspaces by ${\cal H}$.

The UV degree of $V$ w.r.t.~$H$ is just the $\lambda \to \infty$
scaling exponent of $V$ as $u \to \lambda u$
and $a \to a/\lambda$.  First we define:
\beq
V(u,v,q;m,a) \equiv V(k(u,v,q),q;m,a),
\label{Veq}
\eeq
using the fact that the line momenta in \myref{uvdf}
form a basis w.r.t.~$k$.  Then, as $\lambda \to \infty$ we
extract the leading exponent:\footnote{Reisz adds a ``hat''
to the subscript of the degree operator in order to
distinguish it from ``the old polynomial degree.''}
\beq
V(\lambda u,v,q;m,a/\lambda) = \lambda^\nu A(u,v,q;m,a)
+ \ord{\lambda^{\nu-1}} \quad \Leftrightarrow \quad
\degr{u} V = \nu.
\eeq
The UV degree of $C(k,q;m,a)$ is defined similarly.
Combining the two, we have
\beq
\degr{H} \hat I = 4d + \degr{u} V - \degr{u} C,
\label{guvd}
\eeq
where we recall that there are $d$ momenta
internal to $H$.

\subsection{Reisz's theorem}
Convergence of the Feynman integral is then
proven, provided
\beq
\degr{H} \hat I < 0, \quad \forall \quad H \in {\cal H}.
\eeq
In this case, one obtains the remarkable
result:
\beq
\lim_{a \to 0} \hat I = \int_{-\infty}^\infty d^{4L}k
~ \frac{P(k,q;m)}{E(k,q;m)},
\label{fdgc}
\eeq
where
\beq
P(k,q;m) = \lim_{a\to 0} V(k,q;m,a), \quad
E(k,q;m) = \lim_{a\to 0} C(k,q;m,a)
\eeq
are just the continuum limits of the numerator and
denominator resp.  Next I consider the conditions
that are assumed to hold in the course of
proving this result.

\subsection{The \RL\ conditions}
\label{RLc}
In the proof of \myref{guvd} and its consequences,
Reisz makes some assumptions about the Feynman
integrand.  The lattice \PCThm\ of Reisz has
been reviewed by L\"uscher \cite{Luscher:1988sd}, 
and I will make use of his enumeration of
the conditions that are assumed in
the course of the proof.  I refer to these as the {\it \RL\ conditions.} 

First, $V$ satisfies:

\noindent 
\bfe{V1.} {\it There is an integer $\omega$
and function $F$ such that
\beq
V(k,q;m,a) = a^{-\omega} F(ka,qa;ma).
\label{v1e}
\eeq
Moreover, $F$ is smooth, $2\pi$-periodic in the momenta $ka$,
and a polynomial in the masses $ma$.}

\noindent
\bfe{V2.} {\it $V$ has a continuum limit, in
the sense that
\beq
P(k,q;m) = \lim_{a\to 0} V(k,q;m,a)
\eeq
exists.}

L\"uscher notes that ${\bf V1-V2}$ are ``not
very restrictive.''  We will find that they
are satisfied for SF-QCD.

As stated above,
the denominator function $C$ that appears in
\myref{fdgf} is a product of
the denominators of propagators, $C_1,\ldots,C_I$:
\beq
C(k,q;m,a) = \prod_{i=1}^I C_i(\ell_i;m,a).
\label{dend}
\eeq
Here, each $C_i$ depends on a {\it line momentum,} 
described in more detail below.
In SF-QCD, $C_i$ is a trigonometric
function of the line momentum $\ell_i$.

Reisz requires that the line momenta be {\it natural,}
Defn.~3.1 in \cReisz.  That is:\footnote{The asterisk
denotes that I have modified L\"uscher's
statement of the condition in order to bring it
into harmony with stricter definition given by Reisz.}

\noindent
\bfe{L1*.}
{\it The line momenta are of the form
\beq
\ell_i(k,q) = \sum_{j=1}^L C_{ij} k_j + \sum_{\ell=1}^E D_{i\ell} q_\ell
\equiv K_i(k) + Q_i(q).
\label{Lmdf}
\eeq
Moreover, it is assumed that $C_{ij} \in \Zbf, \; D_{i\ell} \in \Rbf$, 
and
\beq
\text{\rm rank} ~ C_{ij} = L, \quad
(C_{i1},\ldots,C_{iL}) \not= 0 \quad \forall \quad i=1,\ldots,I.
\eeq
}

\noindent
\bfe{L2.}
{\it Define the set
\beq
\Lcal = \{ k_1,\ldots,k_L,K_1,\ldots,K_I \},
\eeq
where $K_i=\sum_j C_{ij} k_j$ were defined in \myref{Lmdf}.
Suppose $u_1,\ldots,u_L$ are linearly independent
elements contained in $\Lcal$.  Then the loop momenta
can be expressed in terms of them with integer
coefficients:
\beq
k_i = \sum_{j=1}^L A_{ij} u_j, \quad A_{ij} \in \Zbf.
\eeq
}

Note that this property was used above in \myref{Veq}.
It is in this sense that the line momenta appearing in
\myref{uvdf} form a basis w.r.t.~$k_1,\ldots,k_L$.

The following requirements are imposed on the functions $C_i$
that appear in \myref{dend}:

\noindent
\bfe{C1.}  {\it The propagator denominators can be expressed as
\beq
C_i(\ell_i;m,a) = a^{-2} G_i(a\ell_i;am),
\label{gdfq}
\eeq
where $G_i$ is a smooth function that is $2\pi$-periodic
in the momentum $\ell_i a$.  Also,
$G_i$ is a polynomial in the mass $ma$.}

\noindent
\bfe{C2.}  
{\it The denominators have the conventional
continuum limit:
\beq
\lim_{a \to 0} C_i(\ell_i;m,a) = \ell_i^2 + m_i^2,
\eeq
for fixed $\ell_i,m_i$.
Here, $m_i^2$ is a homogeneous quadratic
polynomial in the mass parameters of the theory.}

\noindent
\bfe{C3.}  {\it There exists an $a_0 > 0$ and an $A > 0$
such that
\beq
|C_i(\ell_i;m,a)| \geq A \( \hat \ell_i^2 + m_i^2 \), \quad
\hat \ell_i^2 \equiv \sum_\mu \frac{4}{a^2} \sin^2(\ell_{i\mu}a/2),
\label{c3e}
\eeq
for all $a \leq a_0$ and all $\ell_i \in \Bcal_a$.}

In Reisz's proof, he does not
require {\bf C3,} but instead a condition that
has the same effect---a lower bound on the
propagator denominator:

\noindent
\bfe{C3*.}
{\it There exists an $a_0 > 0$ and an $A > 0$
such that
\beq
|C_i(\ell_i;m,a)| \geq A \( \ell_i^2 + m_i^2 \),
\eeq
for all $a \leq a_0$ and all $\ell_i \in \Bcal_a$.}

Note that the r.h.s.~of the inequality is a
continuum expression.  The reason that this
is equivalent to {\bf C3} is that $C \equiv \hat \ell_i^2 + m_i^2$ itself
satisfies {\bf C3*.}  Thus we can always
replace the bound in {\bf C3} by the
continuum (rational) expression in {\bf C3*.}
In fact, the essence of Reisz's proof is to replace lattice
expressions by bounds that are rational and
have a continuum interpretation.  For this
reason I prefer {\bf C3*.}


\mys{The conditions of Reisz's theorem vs.~staggered
fermions}
\label{s:pca}
Here I examine the \RL\ conditions in relation to SFs.
It will turn out that the \RL\ conditions
fail in both the 1CB and the MSTB.
The essential reason is a mismatch between
the K\"ahler-Dirac lattice $2a\Zbf^4$ and the {\it gauge lattice}
$a \Zbf^4$.  (The former is the natural lattice
on which to formulate free SFs, whereas the
latter is the lattice on which the pure YM theory
is formulated.)

In these considerations, it is implied that
the $\delta$ functions that appear in the Feyman
rules of \S\S\ref{1cbd}-\ref{mstb} have been integrated
against (except for the overall $\delta$
function that always occurs), leading to the line
momenta $\ell_i(k,q)$.

\subsection{1-component basis}
Consider the denominators of the propagators 
$D_{\mu \nu}^{cd}(k,\ell)$ and $S^{ij}(p,q)$
in relation to the \RL\ conditions.  For the
gluon (B) and quark (F),
\beq
C_B = a^{-2} 
\[ \sum_\mu 4 \sin^2 \frac{\ell_{\mu} a}{2} + (\lambda a)^2\],
\quad
C_F = a^{-2} \[\sum_\mu \sin^2 (\ell_{\mu} a) + (ma)^2\],
\label{1cde}
\eeq
where $\ell$ is the line momentum flowing
into the propagator.  Both
are of the form \myref{gdfq}
and are periodic on $\Bcal_a$; i.e.,
unchanged under $\ell \to
\ell + (2 \pi/a)$.  Thus {\bf C1} is satisfied.
It is also obvious that {\bf C2} holds.
Whereas {\bf C3} holds for $C_B$, it
does not hold for $C_F$.
This is because the latter has lattice poles
away from the origin of $\Bcal_a$, as I now show.

The proof consists of showing that there exist $\ell \in \Bcal_a$
such that {\bf C3} fails.  In particular,
suppose that $\ell = (\pi/a)B$, where
$B \in \Kcal^* \equiv \Kcal-(0^4)$.
Then $C_F = m^2$ and $\hat \ell^2 = 4 a^{-2} \sum_\mu B_\mu
\equiv 4 a^{-2} |B|$, where $\hat \ell^2$ was defined
in \myref{c3e}.  Let $A$ be any strictly positive
real number.  To satisfy
{\bf C3} it is necessary that for sufficiently
small $a_0$, and any $a < a_0$
\beq
m^2 - A(4 a^{-2} |B| + m^2) \geq 0.
\label{mae0}
\eeq
This can only be true for $A<1$.
But, for any $a$ such that
\beq
0 \leq a < a' \equiv \[\frac{4A|B|}{(1-A)m^2}\]^{1/2},
\eeq
condition \myref{mae0} is violated.  Thus we can never
choose $a_0$ small enough to satisfy {\bf C3.}
Put simply, near one of the extra lattice poles, 
$C_F = \ord{m^2}$
whereas $A(\hat \ell^2 + m^2) = A \ord{a^{-2}}$; so, for
small enough $a$ the latter is always larger.  

The numerator of the quark propagator \myref{fp1b}
will contribute to the Feynman numerator $V$ in
\myref{fdgf}, and it is easy to see that it 
satifies {\bf V1-V2.}  The $\delb_a$-functions
that appear are periodic on $\Bcal_a$ by construction.
It is always possible to choose loop momenta $k_i$ 
and line momenta $\ell_i$ such
that $p \equiv \ell_i(k,q)$ in the denominator of \myref{fp1b},
where $q$ is external momentum passing through
the propagator.  Due to the $\pi/a$ violations
of momentum in the 1CB, it is not guaranteed 
that the $\ell_i$ are natural; it is therefore
possible that {\bf L1}-{\bf L2} are also not satisfied.
In any case $\ell_i = C_{ij} k_j + Q_i(q) + z_i (\pi/a)$
with $C_{ij} \in \Zbf$ and $z_i \in \Zbf^4$.  
Then under $k_j \to k_j + (2\pi/a)\nuhat$,
for any loop momentum $k_j$,
the numerator term $\sum_\mu \sin \ell_{i\mu} a$ is unchanged.
The vertex $V_\mu^{c;ij}(k;p,q)$ also
satisfies {\bf V1,} because it is has been
constructed to be periodic on $\Bcal_a$.
It is obvious that the vertex satisfies {\bf V2.}

Thus we see that the \RL\ conditions fail
to hold in the 1CB principally for the
reason that has been pointed out by L\"uscher \cite{Luscher:1988sd}:
the fermion propagator has too many poles in $\Bcala$.
Also worrisome is the $\pi/a$ violations of
momentum.  The latter problem will be eliminated
in the basis that I discuss next.

\subsection{Momentum space taste basis}
\label{mbpc}
Note that the denominators of the MSTB propagators are the
same as in the 1CB.  One has exactly the same violation of
the \RL\ conditions as in the 1CB, 
due to additional quark poles in $\Bcala$.

One might think to instead apply the
\RL\ conditions on the reduced Brillouin
zone $\Bcalaa$, defined in \myref{rbaa}.
This effectively replaces the lattice
spacing $a$ by $b \equiv 2a$.  Whereas
{\bf C3} is satisfied if $\Bcala$ is
replaced by $\Bcal_b \equiv \Bcalaa$, the Feynman rules
do not enjoy $2\pi/b = \pi/a$ periodicity.
Thus {\bf V1} and {\bf C1} would be violated
if we took this approach.
In fact, in the generalization that is introduced
in \S\ref{rcpg} below, the lack of $\pi/a$-periodicity
will be addressed ``head-on''.  

\subsection{The mismatch}
\label{mism}
The problem with $\pi/a$-periodicity is an inevitable
consequence of the mismatch between the K\"ahler-Dirac
lattice and the gauge lattice.  To see this, note
that the Fourier transform \myref{1cft} has been
formulated w.r.t.~the translation invariance group of the gauge
lattice, generated by shifts $\muhat a$.  As a consequence,
the fields $\tilde \phi(k)$ are periodic on the
reciprocal lattice $(2\pi/a)\Zbf^4$.
If not for the gauge fields, we could perform a Fourier transform
w.r.t.~the translation invariance group of the K\"ahler-Dirac lattice,
generated by shifts $\muhat b = 2 \muhat a$.  To accomplish
this, we pass to the position space hypercube basis 
\cite{Gliozzi:1982ib,Kluberg-Stern:1983dg,Patel:1992vu}
before taking the Fourier transform:
\beq
&& \chi(r=2y+A) \equiv \chi_A(y), \quad y\in \Zbf^4, \; A \in \Kcal; \nnn
&& \chi_A(y) = \int_{\Bcal_b} \frac{d^4k}{(2\pi)^4} 
e^{iby \cdot k} \tilde \chi_A(k) =
\int_\Bcalaa \frac{d^4k}{(2\pi)^4} 
e^{i 2a y \cdot k} \tilde \chi_A(k).
\eeq
It is easy to see that the quark propagator
for the $\tilde \chi_A(k)$ fields is $\pi/a$-periodic.
This just follows from the inverse Fourier transform:
\beq
\tilde \chi_A(k) = b^4 \sum_{y \in \Zbf^4}
e^{-ib y \cdot k} \chi_A(y),
\eeq
which clearly leads to 
\beq
\tilde \chi_A(k + (\pi/a)z) = \tilde \chi_A(k) \quad
\forall \quad z \in \Zbf^4.
\eeq
Thus the Feynman rules for $\tilde \chi_A(k + (\pi/a)z)$
and $\tilde \chi_A(k)$ will be identical.  This is
just to say that the reciprocal lattice of
the K\"ahler-Dirac lattice is $(2\pi/b) \Zbf^4 = (\pi/a) \Zbf^4$.

\mys{The Reisz proof generalized}
\label{rcpg}
I now extend the Reisz \PCThm\ and proof such
that the deviations from the \RL\ conditions
can be overcome.  The trick is to use:
\ben
\item[(i)]
the MSTB rules with momentum-conserving
$\delta$-functions, 
\item[(ii)]
integration extended to full periods, 
\item[(iii)]
resolution of
identity on the reduced reciprocal lattice,
and 
\item[(iv)] tranformed Feynman rules that absorb loop
momentum shifts
\beq
k_i \to k_i + (\pi/a) A_i, \quad A_i \in \Kcal.
\label{bbs}
\eeq
\een

The Feynman integral is written in the MSTB.
The domain of integration is $\Bcala$,
using the trick of \S\ref{ddt}.  The
vertices and propagators conserve momentum,
using the equivalence of \S\ref{smsr}.  Of course,
one first integrates and sums against all $\delta$ functions.
This should be done in such a way as to conserve
momentum as it flows through the diagram, using
the $2\pi/a$-periodicity wherever necessary.\footnote{This is
optional because the MSTB Feynman rules only conserve
momentum mod $2\pi/a$.}
Then we are left with just integrals over
loop momenta, which can be routed such
that the line momenta are {\it natural.}
That is, the momenta are routed just as
in a continuum Feynman integral.

\subsection{The denominator}
I now describe the modification that overcomes
the principal difficulties posed by SFs.
One breaks up the line momenta into those corresponding
to bosons (gluons) and fermions (quarks):
$\ell_1^B,\ldots,\ell_{I_B}^B$ and $\ell_1^F,\ldots,\ell_{I_F}^F$
resp.
Then one inserts into the Feynman integral, for each $\ell_i$,
the resolutions of identity that are described in \S\ref{rid1}: 
$1_B(\ell_i^B)$ defined in \myref{bres}
and $1_F(\ell_j^F)$ defined in \myref{fres}.  One obtains
an expression analogous to Reisz's Eq.~(4.4) \cReisz---a 
sum of integrals that comprises a domain decomposition:
\beq
&& \hat I = \sum_{J_B,J_F} \sum_{z_B,z_F}
\hat I(J_B,J_F,z_B,z_F) \equiv \sum_{Jz} \hat I_{Jz}, \nnn
&& J_B \subseteq \{ 1,\ldots,I_B\}, \quad J_F 
\subseteq \{ 1,\ldots,I_F\}, \nnn
&& z_B = (z_{Bi} | i \in J_B), \quad
z_F = (z_{Fj} | j \in J_F),
\eeq
with individual terms of the form:
\beq
\hat I_{Jz} &=& \int_{\Bcala^L} 
\frac{d^{4L}k ~~ V(k,q;m,a)}{\prod_{i=1}^{I_B}
C_B(\ell_i^B;\lambda,a) \prod_{j=1}^{I_F} C_F(\ell_j^F;m,a)} \nnn
&& \times \prod_{i \in J_B} \Theta\(\frac{\pi}{a}\e 
- ||\ell_i^B-\frac{2\pi}{a}z_{Bi}||\) \prod_{i \not\in J_B}
\Theta_\e^B(\ell_i^B) \nnn
&& \times \prod_{j \in J_F} \Theta\(\frac{\pi}{a}\e 
- ||\ell_j^F-\frac{\pi}{a}z_{Fj}||\)
\prod_{j \not\in J_F} \Theta_\e^F(\ell_j^F).
\label{iha}
\eeq
Note that $J$ collectively denotes $J_B,J_F$, and so on.
The decomposition has the following intuitive meaning: 
$\ell_i \in J$ are ``$\e$-near'' to a lattice
pole, whereas $\ell_i \not\in J$ are ``$\e$-far'' from
a lattice pole.

For $\e,a$ sufficiently small, the arguments of Reisz's Appendix D 
\cReisz\ extend in an obvious way
to show that there exists $k^{(0)} \equiv(k_1^{(0)},\ldots,k_L^{(0)})
\in \Bcala^L$ s.t.:
\beq
K_i^B(k^{(0)})=\frac{2\pi}{a} z_{Bi}, \quad
K_j^F(k^{(0)})=\frac{\pi}{a} z_{Fj}, \quad i\in J_B, \quad
j \in J_F.
\eeq
Note that $K_i(k)$ was defined in \myref{Lmdf}.
Using the fact that $\ell_i$ are natural,
it is a trivial extension of Reisz's Lemma D.2 \cReisz\
to prove that there exist reduced reciprocal lattice
vectors
\beq
\Delta_1,\ldots,\Delta_L \in \pia \Zbf^4
\eeq
such that for $i \in J_B, \; j \in J_F$
\beq
\quad K_i^B(\Delta)=\frac{2\pi}{a}z_{Bi}, \quad
K_j^F(\Delta)=\frac{\pi}{a}z_{Fj}.
\eeq
The $\Delta_i$ are determined in terms of a
basis chosen from $\{ K_i^B,K_j^F \}$, as explained
in Reisz's Appendix D \cReisz. 
Thus we define new loop momenta $k'_i$ through:
\beq
k_i = k'_i + \Delta_i \equiv k'_i + \pia \delta_i \quad
\forall \quad i=1,\ldots,L,
\label{yesh}
\eeq
where in the last step integer-valued 4-vectors $\delta_i$
have been introduced for future convenience, following Reisz.
A new domain of integration results:
\beq
\s_J = \left\{ k' \in \Rbf^{4L} ~ \Big| ~ -\pia - \Delta_{i\mu} < k'_{i\mu} \leq
\pia - \Delta_{i\mu} \right\},
\label{sdom}
\eeq
identical to Reisz's Eq.~(4.7) \cReisz.

For the line momenta $\ell_i^B \in J_B, 
\ell_j^F \in J_F$, \myref{yesh} has the effect
\beq
\ell_i^B(k) = \ell_i^B(k') + \frac{2\pi}{a} z_i^B, \quad
\ell_j^F(k) = \ell_j^F(k') + \frac{\pi}{a} z_j^F,
\eeq
a generalization of Reisz's (4.5) \cReisz.
When this is accounted for in \myref{iha}, the
Heaviside step functions in \myref{iha} just force $\ell_i^B(k') \in J_B, \; 
\ell_j^F(k') \in J_F$ into the $\e$-neighborhood of the (unique) pole in
$\Bcala$ and $\Bcalaa$ respectively.  
As a consequence the following bounds hold:
\beq
&& C_B^{-1}(\ell_i^B \in J_B) \leq \alpha_B (\ell_i^B(k')^2+\lambda^2)^{-1}, \nnn
&& C_F^{-1}(\ell_j^F \in J_F) \leq \alpha_F (\ell_j^F(k')^2+m^2)^{-1},
\eeq
generalizations of Reisz's (4.8) \cReisz.
Here, $\alpha_B,\alpha_F$ are constants that
always exist for $\e,a$ sufficiently small.
For $\ell_i \not\in J$, the line momenta
are outside of the balls of radius $\e \pi/a$
that are centered on sites of the (reduced)
reciprocal lattice for (quarks) gluons.
Therefore they are bounded by:
\beq
C_B^{-1}(\ell_i^B \not\in J_B) \leq \gamma_B a^2, \quad
C_F^{-1}(\ell_j^F \not\in J_F) \leq \gamma_F a^2,
\eeq
generalizations of Reisz's (4.9) \cReisz.
Here, $\gamma_B,\gamma_F$ are constants that
always exist for $\e,a$ sufficiently small.

For the line momenta $\ell_i^B \not\in J_B, 
\ell_j^F \not\in J_F$, the shift \myref{yesh} is only
guaranteed to have
\beq
\ell_i^B(k) - \ell_i^B(k') = C^B_{im} \Delta_m \in \pia \Zbf^4, \quad
\ell_j^F(k) - \ell_j^F(k') = C^F_{jm} \Delta_m \in \pia \Zbf^4.
\eeq
Whereas $\Theta_\e^F$ is $\pi/a$-periodic,
the function $\Theta_\e^B$ is only $2\pi/a$-periodic.
Some explicit dependence on $\Delta$ will result,
and will be addressed below.
Gathering together the various results,
we can bound \myref{iha} by:
\beq
\hat I_{Jz} &\leq& 
\alpha_B^{h_B} \alpha_F^{h_F} (\gamma_Ba^2)^{(I_B-h_B)} 
(\gamma_F a^2)^{(I_F-h_F)} \nnn
&& \quad \times \int_{\s_J} 
d^{4L}k' ~~ V(k'+\Delta,q;m,a) \nnn
&& \quad \times \prod_{i \in J_B} \[ (\ell^B_i(k')^2+\lambda^2)^{-1}
\Theta\( \frac{\pi}{a} \e - ||\ell_i^B(k')|| \) \] \nnn
&& \quad \times \prod_{j \in J_F} \[ (\ell_j^F(k')^2+m^2)^{-1}
\Theta\( \frac{\pi}{a} \e - ||\ell_j^F(k')|| \) \] \nnn
&& \quad \times \prod_{i \not\in J_B} \Theta_\e^B(\ell_i^B(k')+C^B_{im} \Delta_m)
\prod_{j \not\in J_F} \Theta_\e^F(\ell_j^F(k')).
\label{myf}
\eeq
Here, the $h_B,h_F$ are the number of elements
in $J_B,J_F$ resp.; i.e., the number of
line momenta that are $\e$-near to lattice poles.
The denominator has been expressed entirely in terms
of rational functions.  
The numerator $V$ requires further study:
the shifted loop momentum argument $k'+\Delta$ 
can be accomodated into Reisz's techniques
to bound the numerator, as will be discussed further
in \S\ref{nuse} and \S\ref{urp} below.
It will be seen that
the $\Theta,\Theta_\e^B, \Theta_\e^F$ functions
do not pose any difficulty, as they 
just restrict the domain of loop integration.
With the bound in the form \myref{myf}, it is
quite simple to extend the remainder of
Reisz's manipulations.  Using
them, I will formulate and prove the SF \PCThm.


\subsection{The numerator}
\label{nuse}
The magnitude of the SF numerator is also easy to
estimate, using the decomposition $J_F$ and the
shifted line momenta $\ell_j(k')$.  However,
cancellations associated with the spin-taste
algebra will be important to take into account
in order to get the correct UV degree for a given
diagram.  For this reason it is better to
abide by Reisz's approach and treat the numerator
$V$ as a whole.

I now make a few remarks regarding the effect
of the shift $\Delta$ that appears in the
numerator of \myref{myf}.  This will lead to
modifications of propagator numerators and of
vertex factors.  In the words of Reisz,
a generic shift $\Delta_i \in \Rbf^4$ ``would
produce explicit negative powers in the lattice
spacing destroying convergence.''  However, though
the shift involved here is not an invariance
of the Feynman integrand, it is nevertheless special:  $\Delta_i \in
(\pi/a) \Zbf^4$.  As I now discuss, it is possible
to eliminate this explicit $a^{-1}$ through a
transformation in the form of the Feynman integrand.
The transformed integrand
trades $\sin \leftrightarrow \cos$ in various places,
and/or introduces factors of $(-1)$.  Furthermore,
the number of possibilities for how the
propagators and vertex factors are transformed is 
finite.  The UV degree of the transformed numerator
is then determined in accordance with Reisz's definition.
This degree is then used in a generalized computation
of the UV degree of the integral $\hat I$, as
will be seen in \S\ref{gth} below.

As an example of the transformation induced
by $\Delta$, consider the quark-gluon vertex
\myref{ms1v}.  For the sake of argument,
suppose that each line entering the vertex
is internal, with
\beq
&& p \to \ell_1, \quad q \to \ell_2, \quad k \to \ell_3, \nnn
&& \ell_i = C_{ij}k_j + Q_i(q) \quad \forall \quad i=1,2,3.
\label{p2L}
\eeq
Thus in the redefinition \myref{yesh}
the momentum-dependent factor in the vertex transforms as:
\beq
&& e^{ia\ell_{1\mu}(k)} + e^{-ia(\ell_{1\mu}(k) +
\ell_{3\mu}(k))} = \nnn
&& \qquad (-)^{ C_{1i} \delta_{i\mu} }
\[ e^{ia\ell_{1\mu}(k')} + (-)^{ C_{3i} \delta_{i\mu} }
e^{-ia(\ell_{1\mu}(k') + \ell_{3\mu}(k'))} \].
\label{trve}
\eeq
Note that since the
line momenta are natural, $C_{ij} \in \Zbf$.  It
follows that, as promised, factors of $(-1)$ have
been introduced.  Momentum conservation implies
\beq
C_{1i} + C_{2i} + C_{3i} = 0,
\eeq
but this still allows for the transformation \myref{trve}
to have a notrivial effect on the quark-gluon vertex.
In the case that $C_{3i} \delta_{i\mu} = 1 \mod 2$,
the factor $\cos(\ell_{1\mu}(k)a
+ (1/2) \ell_{3\mu}(k)a)$ is exchanged for $\sin(\ell_{1\mu}(k')a
+ (1/2) \ell_{3\mu}(k')a)$.  [Here, it is implicit
that overall exponentials are factored out
to rewrite the expression using trigonometric functions.]
Since the latter starts at
$\ord{a}$, rather than $\ord{1}$, the UV properties of
the vertex are changed in a significant way (lowered by 1).
This does not destroy the convergence of the numerator;
in fact, it improves it.

As another example, consider the triple-gluon
vertex.  In the conventions taken here,\footnote{See
for example Eq.~(3.225) of \cite{Montvay}, taking into
account the $2\pi/a$-periodic Fourier transform
convention \myref{1cft} that I use.}
\beq
V_{\nu \rho \mu}^{abc}(p,q,k) &=& \frac{g}{4} f^{abc}
\delta_{\mu \nu} \delba(p+q+k) \( e^{iak_\rho} -
e^{-ia(k_\rho+q_\rho)} \) \( 1 + e^{-ia(p_\mu+k_\mu)} \) \nnn
&& \qquad + ~ \text{cyclic permutations}.
\eeq
Applying \myref{p2L}, one finds:
\beq
&& \( e^{ ia \ell_{3\rho}(k) } 
- e^{-ia( \ell_{3\rho}(k) + \ell_{2\rho}(k) ) } \) 
\( 1 + e^{-ia( \ell_{1\mu}(k)+\ell_{3\mu}(k) )} \)
= \nnn
&& \qquad (-)^{ C_{3i} \delta_{i\rho} } 
\( e^{ ia \ell_{3\rho}(k') } 
- (-)^{ C_{2i} \delta_{i\rho} } 
e^{ -ia( \ell_{3\rho}(k') + \ell_{2\rho}(k') ) } \) \nnn
&& \qquad \times ~
\( 1 + (-)^{(C_{1i}+C_{2i})\delta_{i\mu}}
e^{-ia(\ell_{1\mu}(k')+\ell_{3\mu}(k')) } \).
\label{vttr}
\eeq
Here again, the explicit $a^{-1}$ contained in $\Delta_i$
is traded for factors of $(-1)$, that in some cases
interchange $\sin \leftrightarrow \cos$.  Raising
or lowering the UV degree by 1.

An increase in the UV degree of the numerator under
transformations such as \myref{vttr} will not ``destroy''
the ``convergence'' of the Feynman integral.  Rather,
it will only make manifest the cutoff dependence.
To the extent that a Feynman integral has positive
UV degree, subtractions are required in order to
have a convergent result, regardless of the basis
of loop momenta.

Quite generally, the factors in the numerator $V(k,q;m,a)$
are trigonometric functions that are $2\pi/a$-periodic.
The redefinition $k_i=k'_i+(\pi/a)\delta_i$, $\delta_i \in \Zbf^4$
always results in a half- or full-period shift.
The rule is that, prior to computing the UV degree, 
one should eliminate the explicit
$\pi/a$ factor using elementary trigonometric
identities, as has just been illustrated for the
quark-gluon and triple-gluon vertices.  

The best strategy to deal with this is to extend
the Feynman rules to incorporate $\pi/a$ shifts.
Then the integral \myref{myf} should be interpreted
in terms of these new rules.  Symbolically,
\beq
&& V(k'+(\pi/a)\delta,q;m,a) \equiv V(k'+(\pi/a)A,q;m,a)
\equiv V_A(k',q;m,a), \nnn
&& A_{i\mu} = \delta_{i\mu} \mod 2,
\quad A_i \in \Kcal.
\label{vtrA}
\eeq
Here, $V_A$ is written in terms of the generalized Feynman
rules.  (Note that in the second step 
the $2\pi/a$-periodicity has been used to express
the numerator $V$ in terms of the transformed one
with an index restricted to $A \in \Kcal^L$.)
Once this has been done, all of Reisz's techniques
for the UV degree analysis of the numerator apply.
This can be seen from the fact that the numerator
satisfies the \RL\ conditions, after the explicit
factors of $\pi/a$ have been eliminated.  The essential
reason for this is that sine and cosine are analytic
functions.

\subsection{The generalized theorem}
\label{gth}
These considerations lead to the following 
generalization of Reisz's theorem:

\noindent
\ite{Definition.}
Let $F_A= V_A/C_A, \; A \in \Kcal^L$ denote the transformed
Feynman integrand.  That is:
\beq
&& V_A(k,q;m,a) = V(k+(\pi/a)A,q;m,a), \nnn
&& C_A(k,q;m,a) = C(k+(\pi/a)A,q;m,a).
\eeq
Generalize the UV degree as follows:
\beq
&& \ddegr{u} F = \max_{A \in \Kcal^L} \degr{u} F_A,
\quad \degr{u} F_A = \degr{u} V_A - \degr{u} C_A; \nnn
&& \ddegr{H} \hat I = 4d + \ddegr{u} F.
\eeq
Recall that $u_1,\ldots,u_d$ parameterizes
the Zimmermann subspace $H$.

\noindent
\ite{Proposition.}
Suppose that
\beq
\ddegr{H} \hat I < 0 \quad \forall \quad H \in {\cal H}.
\label{Idn}
\eeq
Then $\hat I$ converges, and
\beq
\lim_{a\to 0} \hat I = \sum_{A \in \Kcal^L}
\int_{-\infty}^\infty d^{4L}k ~ \frac{P_A(k,q;m)}{E_A(k,q;m)},
\label{clim}
\eeq
where
\beq
P_A(k,q;m) = \lim_{a\to 0} V_A(k,q;m,a), \quad
E_A(k,q;m) = \lim_{a\to 0} C_A(k,q;m,a)
\eeq
are just the continuum limits of the numerator and
denominator resp.  This indicates that various
regions of loop momenta may contribute to
the continuum limit, due to the presence of
doublers in the fermion spectrum.

\subsection{Proof}
\label{urp}
Starting with \myref{iha}, one makes the redefinition \myref{yesh}.
Then the numerator is replaced by $V_A(k',q;m,a)$, as in
\myref{vtrA}.  Once this has been done, $\hat I_{Jz}$
is in the form considered by Reisz.  Due to the
assumption \myref{Idn}, the remainder $R_A$ in the
decomposition
\beq
V_A(k,q;m,a) = P_A(k,q;m) + R_A(k,q;m,a)
\label{VAde}
\eeq
does not contribute in the continuum limit, as
follows from Reisz's arguments in \S7 of \cReisz.
Thus one can replace $V_A$ by the rational function $P_A$
in the numerator of $\hat I_{Jz}$.
Furthermore, Reisz's arguments show that the $\hat I_{Jz}$ term
that maps to the index $A \in \Kcal^L$ just yields
\beq
I_A = \int_{-\infty}^\infty d^{4L}k
~ \frac{P_A(k,q;m)}{E_A(k,q;m)}
\label{IAdf}
\eeq
in the continuum limit.  The result \myref{clim}
follows immediately.

To clarify this, I discuss some of the details of \S7 of Reisz \cReisz.
Taking into account the decomposition \myref{VAde} and
the effect of \myref{yesh} and \myref{vtrA} on \myref{iha},
we have the decomposition
\beq
\hat I_{Jz} = \hat I_{Jz}^{0} + \hat I_{Jz}^{R},
\eeq
where
\beq
\hat I_{Jz}^{0} &=& 
\int_{\s_J} \frac{d^{4L}k'~~ P_A(k',q;m)}{C_A(k',q;m,a)} \nnn
&& \quad \times \prod_{i \in J_B} 
\Theta\( \frac{\pi}{a} \e - ||\ell_i^B(k')|| \)
\prod_{j \in J_F} 
\Theta\( \frac{\pi}{a} \e - ||\ell_j^F(k')|| \) \nnn
&& \quad \times \prod_{i \not\in J_B} \Theta_\e^B(\ell_i^B(k')+C^B_{im} \Delta_m)
\prod_{j \not\in J_F} \Theta_\e^F(\ell_j^F(k')),
\eeq
and $\hat I_{Jz}^{R}$ is defined with $P_A$ replaced by $R_A$.
Note that there is a correspondence between the index $A$
and the index $Jz$, since the latter determines the
shift \myref{yesh} that is required.
When one compares to Reisz's (7.1)-(7.2) of \cReisz,
the only difference that one finds is in the factor
$\prod_{i \not\in J_B} \Theta_\e^B(\ell_i^B(k')+C^B_{im} \Delta_m)$,
with its extra argument $C^B_{im} \Delta_m$.  This only restricts
the domain of integration; it can only make the integral smaller,
so Reisz's bounds still hold.  E.g.,
\beq
|\hat I_{Jz}^{0}| \leq \bar I_{Jz}^{0}
&=& \alpha_B^{h_B} \alpha_F^{h_F} (\gamma_Ba^2)^{(I_B-h_B)} 
(\gamma_F a^2)^{(I_F-h_F)} \nnn && \times \int_{\kappa_J} 
\frac{d^{4L}k' ~~ |P_A(k',q;m)|}{\prod_{i \in J_B}(\ell_i^{B}(k')^2 + \lambda^2)
\prod_{j \in J_F}(\ell_j^{F}(k')^2 + m^2)} ,
\eeq
where $\kappa_J$ is the same domain as in Reisz's (7.6) \cReisz.
(Following Reisz, the domain \myref{sdom} is extended, $\s_J \to \kappa_J$,
which can only increase the value of the bounding continuum integral
$\bar I_{Jz}^0$.)
Similar remarks apply to $\hat I_{Jz}^{R}$.  Note that this
is just an exploitation of the bound \myref{myf}.  

It follows
that the remainder of Reisz's arguments of \S7 \cReisz\
apply, with only the following modification.
It is still true that $\bar I_{Jz}^0$ has a
nonvanishing continuum limit only if all
line momenta are near poles:  $J_B=\{1,\ldots,I_B\}$
and $J_F=\{1,\ldots,I_F\}$.  However,
rather than Reisz's Eq.~(7.18) \cReisz, we must write
($I=I_B+I_F$ and $i$ collectively denotes all
line momenta):
\beq
K_i(k) = \sum_{j=1}^L C_{ij} k_j = \pia z_i, \quad
i=1,\ldots,I,
\label{Kke}
\eeq
which differs in that $2\pi/a$ has been replaced
by $\pi/a$ for the coefficient of $z_i$.  Just as in
his case, this equation has a unique solution,
due to ${\rm rank} (C_{ij}) = m$.  Due to the
naturalness of the line momenta, the solution
is of the form
\beq
k_i = (\pi/a)\delta_i, \quad \delta_i \in \Zbf^4.
\label{imeq}
\eeq
These
are nothing but the $\delta_i$ that are used in
the shift \myref{yesh}.  

Now consider the possible solutions to \myref{imeq},
given $k_i \in \Bcala$.  Unlike what occurs in
the case considered by Reisz, there are
multiple possibilities for $\delta_i$ that will work.  In fact,
they are nothing other than the $A_i \in \Kcal$.
That is, the sum over $A \in \Kcal^L$ that is
taken in \myref{clim} is in one-to-one correspondence
with the set of $k_i = (\pi/a)\delta_i$ that
are solutions to \myref{imeq}.  (As will be
seen in the example of \S\ref{efe}, it can
happen that for certain $A \equiv \delta \in \Kcal^L$,
we get $z_i \not\in 2\Zbf^4$ where $i$ corresponds
to one of the gluon lines, $\ell^B_i$.  In
that case the contribution $I_A$ will vanish,
since the gluon is always far from the pole
in $\bar I_{Jz}^0$.)  From this
we see that the sum in \myref{clim} is just
the modification that is required to extend
the arguments of Reisz's proof.

\subsection{Simple examples}
\subsubsection{1-loop with external gluons}
\begin{figure}
\begin{center}
\includegraphics[width=2.5in,height=2.5in,bb=100 400 500 750,clip]{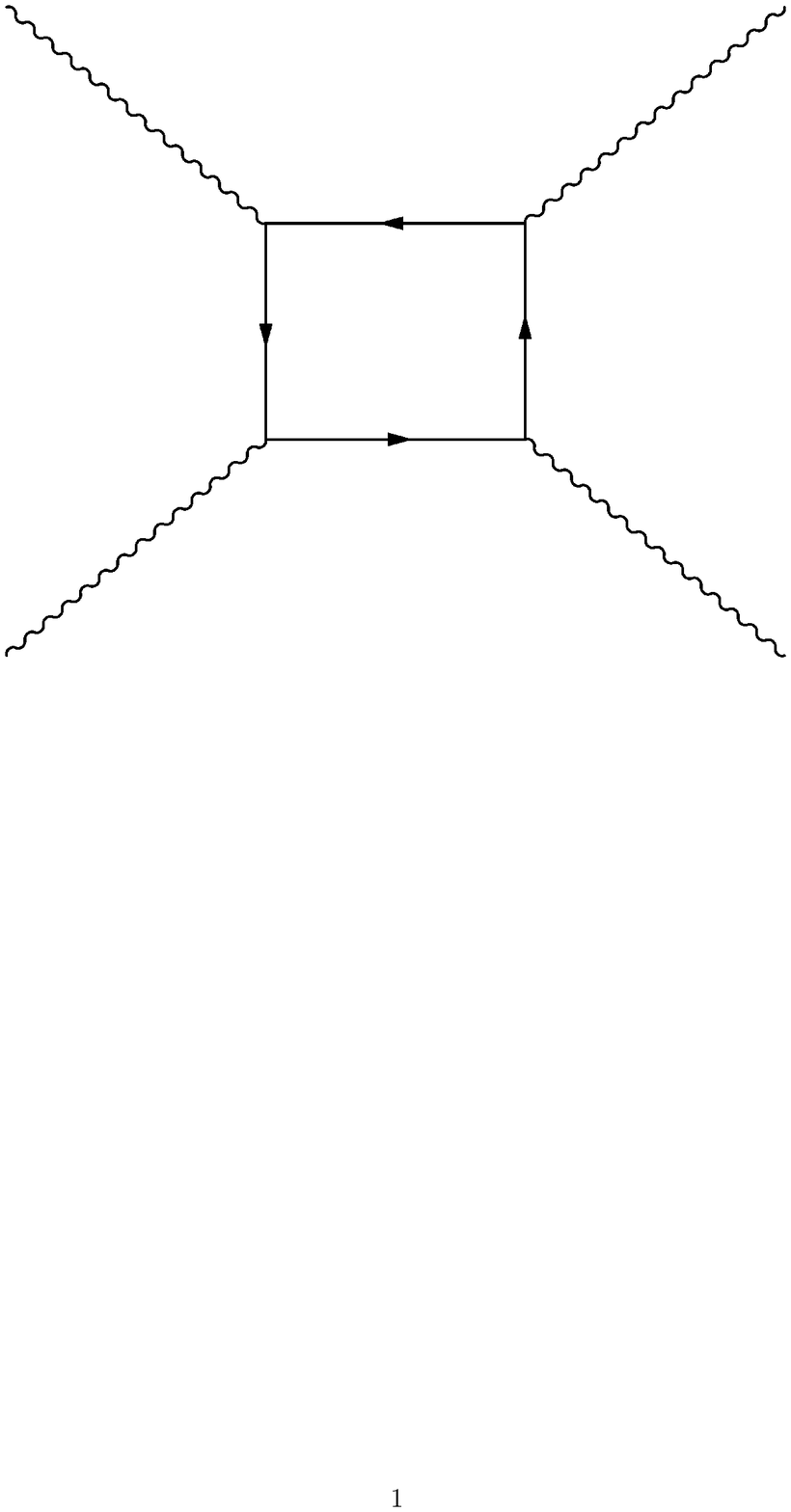}
\caption{Example of finite diagram where all
16 choices of $A \in \Kcal$ contribute.
\label{4glu}}
\end{center}
\end{figure}

Consider any 1-fermion-loop diagram with external
gluon legs.\footnote{Vacuum polarization will of course
require a subtraction for the \PCThm\ to hold.}  For example,
the 4-gluon diagram Fig.~\ref{4glu}.
Recall that in the
MSTB rules, a factor of $1/16$ is supplied for each fermion
loop, due to the extension of the integration domain.
On the other hand, when one takes the continuum limit
of the numerator and denominator, only one pole region ($k\approx0$)
occurs in the continuum Feynman integrand [Eq.~\myref{IAdf} with
$A=(0,0,0,0)$.].  It is only when
we include all 16 contributions that come from the
sum over $A \in \Kcal$ that appears in \myref{clim}
that we get the correct overall factor.

In detail, the denominator of the integrand
is invariant under the shift of
loop momentum, $k \to k + (\pi/a)A$,
since it involves only the SF propagator.  
Taking into account the change in the
SF propagator numerators and the gluon-quark
vertices, it can be shown that the change in the numerator of the
integrand is equivalent to just:
\beq
\msg{\gamma_\mu}{1} \to (-)^{A_\mu} \msg{\gamma_\mu}{1},
\eeq
in every place that $\msg{\gamma_\mu}{1}$ appears,
and everything else left as it was before the
shift in loop momentum.
This is just an automorphism of the Clifford
algebra, and will not change the value of the
character---i.e., the trace that occurs in
the numerator.  Thus each of the 16
Feynman integrands, $A \in \Kcal$, are identical in their
continuum limit to the one obtained
at $A=0$.  This just cancels
the factor of $1/16$ to give the desired result.

\subsubsection{1-loop with external fermions}
\label{efe}
\begin{figure}
\begin{center}
\includegraphics[width=2.5in,height=2.5in,bb=100 400 500 750,clip]{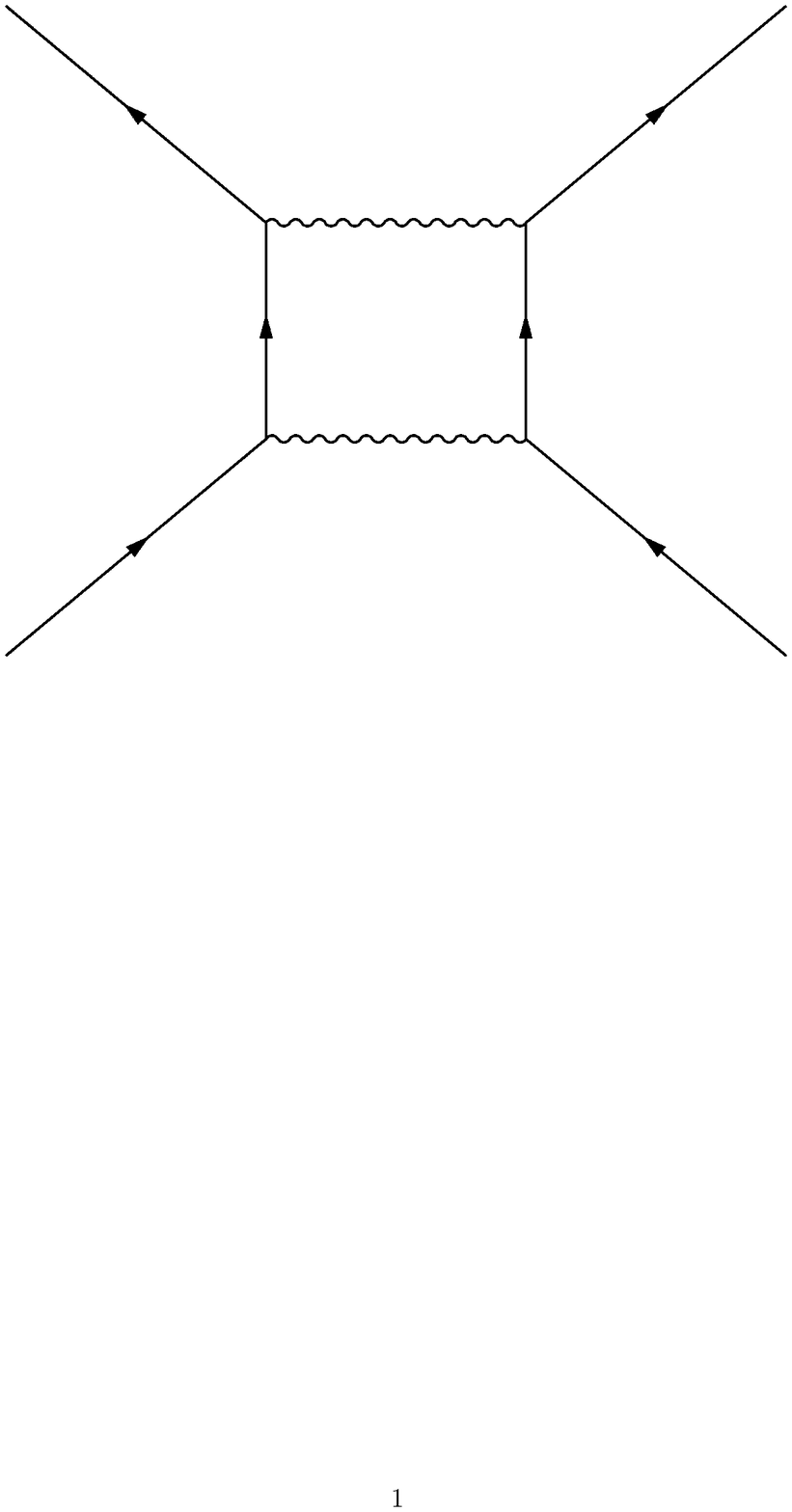}
\caption{Example of finite diagram where only
1 choice of $A \in \Kcal$ contributes.
\label{4fer}}
\end{center}
\end{figure}

Next consider diagrams such as Fig.~\ref{4fer}.
In this case the gluons contribute to the
denominator.  When the shift $k \to k+ (\pi/a) A$
is performed, with $A \not= 0$, the gluon
denominators are moved far from their poles when $k \approx 0$.
It follows that $C_A^{-1} \sim a^4$, where $C_A$
is the transformed denominator of the Feynman
integrand.  The numerator of the integrand
$V_A$ is well-behaved as $a \to 0$.  Thus
for $A \not=0$ the integrand vanishes in
the $a \to 0$ limit.  This is just as required:
there is no $1/16$ factor to be compensated
because we do not have a fermion loop.  Only
the $A=0$ contribution survives in the $a\to 0$
limit, and the result is just the usual continuum
expression.

\mys{Discussion}
\label{s:con}
In this article I have demonstrated how to extend
the techniques of Reisz's power-counting theorem to the case of
staggered fermions.  It is fortunate that with
a few straightforward modifications, the bulk
of Reisz's arguments apply.  It is of great practical importance
that his power-counting theorem for generalized
continuum Feynman integrals (\S5 of \cReisz)
continues to be applicable.
One thing that remains to be done is to use
the staggered fermion \PCThm\ to prove
perturbative renormalizability, following 
\cite{Reisz:1987px,Reisz:1988kk}.  Also, applications of the
theorem to higher orders in perturbation theory should be explored
in more detail.  Generalizations to other sorts of theories
that contain doublers could also be considered,
since the technique that has been
introduced here is not
very specialized.  It is worth emphasizing
that all of the manipulations that
were performed here apply equally well
to improved staggered fermion QCD.

\vspace{30pt}

\noindent
{\bf \Large Acknowledgements}

\vspace{5pt}

\noindent
I benefitted from communications with David Adams.
This work was supported in part by the U.S.~Department of Energy
under grant No.~DE-FG02-94ER-40823.

\myappendix

\mys{Resolutions of identity}
\label{rid1}
\subsection{Resolution on $\Bcal_a$}
In \S4 of \cReisz, Reisz introduces the following step function:
\beq
\Theta_\e^B(\ell) = \left\{ \begin{array}{ll}
0 & \hbox{if $||\ell-\frac{2\pi}{a}z||
< \frac{\pi}{a}\e$ for some $z \in \Zbf^4$,} \\
1 & \hbox{otherwise.}
\end{array} \right.
\eeq
A superscript $B$ has been affixed to distinguish it
from another step function that will be defined below.
For any $\ell$ one can resolve identity as:
\beq
1 = 1_B(\ell) &\equiv& \Theta_\e^B(\ell) + \sum_{z\in \Zbf^4}
\Theta \( \frac{\pi}{a}\e - \bnb \ell-\frac{2\pi}{a}z \bnb \).
\label{bres}
\eeq
Here, $\Theta$ is Heaviside's unit step function.

\subsection{Resolution on $\Bcal_{2a}$}
Define a step-function analogous to Reisz's:
\beq
\Theta_\e^F(\ell) = \left\{ \begin{array}{ll}
0 & \hbox{if $||\ell-\frac{\pi}{a}z||
< \frac{\pi}{a}\e$ for some $z \in \Zbf^4$,} \\
1 & \hbox{otherwise.}
\end{array} \right.
\eeq
For any $\ell$ one can resolve identity as:
\beq
1 = 1_F(\ell) &\equiv& \Theta_\e^F(\ell) + \sum_{z \in \Zbf^4}
\Theta \( \frac{\pi}{a}\e - \bnb \ell-\frac{\pi}{a}z \bnb \).
\label{fres}
\eeq
As above, $\Theta$ is Heaviside's unit step function.
This resolution is useful for line momenta of
SFs, since it isolates the regions that are near
SF poles.

\mys{Simplification of MSTB rules}
\label{smsr}
Here I establish a very important simplification
that is used to derive the MSTB rules \myref{msfp}-\myref{ms1v}.  
It was employed, for instance, by Patel and Sharpe
\cite{Patel:1992vu}.

\subsection{$\delta$-function transformation}
Without loss of generality, for any momenta
$p,q \in \Rbf^4$ we can write
\beq
p=p'+\pia A, \quad q=q'+\pia B, \quad
p',q' \in \Rbf^4, \quad A,B \in \Kcal.
\label{momc}
\eeq
That is, $p',q'$ are not necessarily
in the reduced first Brillouin zone $\Bcalaa$.
Even if $p,q$ are integrated over $\Bcala$,
we can impose that $p',q'$ are also integrated 
over $\Bcala$ by using the equivalence that
is established in \S\ref{ddt}.  In
fact, this is exactly what I do when passing
to the MSTB from the 1CB.  It is also what has
been done by Patel and Sharpe \cite{Patel:1992vu}.

It is not hard to check that the
$\delta$-functions that appear in the
1CB quark propagator \myref{fp1b} and vertex
\myref{1dpr} obey the identities
\beq
&& \delba \(p'+q'+k+\pia (A+B+\mubar)\) = \nnn
&& \prod_{\nu=1}^4 \[ \delba (p'_\nu+q'_\nu+k_\nu) \deltw_{A_\nu + B_\nu
+ \mubar_\nu,0} 
+ \delba \(p'_\nu+q'_\nu+k_\nu + \pia\) \deltw_{A_\nu + B_\nu
+ \mubar_\nu,1} \], \qquad
\label{dfid}
\eeq
with $k=0$ in the quark propagator \myref{fp1b},
and $\deltw$ a Kr\"onecker $\delta$ $\modtxt 2$.
It is worth noting that in the transition from
\myref{fp1b}-\myref{1dpr} to \myref{msfp}-\myref{ms1v},
the identity\footnote{This identity
is easily checked using, for example, the definitions
given in \S2.1-2.2 of \cite{Patel:1992vu}.}
\beq
\msg{\gamma_\mu}{1}_{A,B} =
(-)^{A_\mu} \prod_{\nu=1}^4 \deltw_{A_\nu + B_\nu
+ \mubar_\nu,0} \equiv (-)^{A_\mu} \deltw_{A + B
+ \mubar,0}
\eeq
is used.  The factor $(-)^{A_\mu}$ is obtained from the
$p$-dependent prefactors of the $\delba$-functions
in \myref{fp1b}-\myref{1dpr} under $p=p'+(\pi/a)A$.
However, the other $\delta$-functions that occur
in \myref{dfid} need to be taken into account.
Next I will demonstrate the equivalence that
allows us to eliminate the $\delta$ functions
that violate momentum conservation by $\pi/a$.
In the proof the \PCThm, this is key to obtaining
{\it natural} line momenta in the continuum, 
bounding integrals $\bar I_{Jz}^0$ and $\bar I_{Jz}^R$.

\subsection{Equivalence}
As mentioned above, the trick of \S\ref{ddt} is used
to extend the integration of $p',q'$ to $\Bcala$.
Then a typical integral against the $\delba(\cdots + \pi/a)$
parts of \myref{dfid} takes the form:
\beq
I &=& \frac{1}{(16)^2} \int_{-\pi/a}^{\pi/a} dp'_\nu
\int_{-\pi/a}^{\pi/a} dq'_\nu \int_{-\pi/a}^{\pi/a} dk_\nu
\sum_{A,B} \nnn && \times \delba \( p'_\nu+q'_\nu+k_\nu + \pia \)
\deltw_{A_\nu + B_\nu + \mubar_\nu,1} ~f\( p'_\nu+ \pia A_\nu, 
q'_\nu+\pia B_\nu, k_\nu;\ldots \),
\eeq
where $f$ represents the rest of the integrand, 
where ``$\ldots$'' corresponds to integrals over other momenta components
and other momenta.  I hide integrations
over these other variables, for simplicity of
notation.  The functional form of $f$, namely the
way that $p'$ and $A$ appear together, etc., is
guaranteed by the fact that we start from the 1CB.

Next I make the redefinitions
\beq
q'_\nu \to q'_\nu - \pia, \quad B_\nu \to B_\nu + 1,
\eeq
to obtain
\beq
I &=& \frac{1}{(16)^2} \int_{-\pi/a}^{\pi/a} dp'_\nu
\int_{0}^{2\pi/a} dq'_\nu \int_{-\pi/a}^{\pi/a} dk_\nu
\sum_{A,B} \nnn && \times \delba \( p'_\nu+q'_\nu+k_\nu \)
\deltw_{A_\nu + B_\nu + \mubar_\nu,0} ~f\( p'_\nu+ \pia A_\nu, 
q'_\nu+\pia B_\nu, k_\nu;\ldots \).
\eeq
Finally, since the integration over $q'_\nu$ is a full
period, we can shift the domain
\beq
\int_{0}^{2\pi/a} dq'_\nu \to 
\int_{-\pi/a}^{\pi/a} dq'_\nu
\eeq
at no cost.  This establishes the equivalence $\delba(\cdots + \pi/a)$
to the momentum conserving terms $\delba(\cdots)$.  In essence,
we trade momentum violation by $\pi/a$ for a shift in taste.
Thus we find that under the integration $\int_\Bcala d^4q'$,
\beq
&& \prod_{\nu=1}^4 \[ \delba (p'_\nu+q'_\nu+k_\nu) \deltw_{A_\nu + B_\nu
+ \mubar_\nu,0} 
+ \delba \(p'_\nu+q'_\nu+k_\nu + \pia\) \deltw_{A_\nu + B_\nu
+ \mubar_\nu,1} \] \nnn
&& \simeq 16 \delba ( p' + q' +k) \deltw_{A + B + \mubar,0}.
\eeq
We could also have used the $p'$ integration in
these manipulations.  If 2 momenta are external,
we cannot use this equivalence and more care is
required \cite{Sharpe:1993ur}.  This technicality
does not affect the power-counting considerations
here, because we are only interested in 1PI loop diagrams (the
aim is to study renormalization),
which do not contain vertices with 2 external momenta.

\mys{Domain extension in MSTB}
\label{ddt}
Here I prove that after the decomposition \myref{momc} of
momenta for the transition 1CB $\to$ MSTB, we can double the
domain of integration for $p'$, etc.  The point is to
obtain integration over a full period.  It
also is invoked in the simplification of
\S\ref{smsr}.

In the manipulations of this paragraph, $p$ will denote any
of the components $p_\mu$ of a momentum that is
integrated over in the Feynman rules in the 1CB.
We use equivalence of integration over any full period to
write:
\beq
\int_{-\pi/a}^{\pi/a} dp \simeq
\int_{-\pi/2a}^{3\pi/2a} dp = 
\half \( \int_{-\pi/2a}^{3\pi/2a} dp_1 + \int_{-\pi/2a}^{3\pi/2a} dp_2 \),
\label{un1}
\eeq
where in the last step the integration has been prepared
for further manipulations.  Then we decompose $p_1,p_2$
onto separate reduced domains:
\beq
&& p_1=p'_1+ \pia A_1, \quad p'_1 \in (-\pi/2a,\pi/2a], \quad
A_1 \in \{ 0,1 \}, \nnn
&& p_2=p'_2 - \pia A_2, \quad p'_2 \in (\pi/2a,3\pi/2a], \quad
A_2 \in \{ 0,1 \}.
\eeq
The integration of \myref{un1} can then be expressed 
equivalently as:
\beq
\half \( \int_{-\pi/2a}^{\pi/2a} dp'_1 \sum_{A_1}
+ \int_{\pi/2a}^{3\pi/2a} dp'_2 \sum_{A_2} \)
= \half \sum_A \int_{-\pi/2a}^{3\pi/2a}
\simeq \half \sum_A \int_{-\pi/a}^{\pi/a},
\eeq
where in the last step I have again used
the equivalence of integrations over a full period.

Extending this manipulation to all components, we obtain
the identity
\beq
\int_\Bcala d^4p (\cdots)
= \frac{1}{16} \sum_{A\in\Kcal} \int_\Bcala d^4p' (
\cdots)_{p \equiv p'+\pia A} .
\eeq

\end{document}